\documentstyle[prl,aps,psfig]{revtex}
\begin{document}
\draft
\twocolumn[\hsize\textwidth\columnwidth\hsize\csname
@twocolumnfalse\endcsname

\title{Correlated ground states with (spontaneously) broken time-reversal
symmetry}

\author{Behnam Farid}
\address{Max-Planck-Institut f\"ur Festk\"orperforschung,
Heisenbergstra\ss e 1,\\
70569 Stuttgart, Federal Republic of Germany }

\date{\today}

\maketitle

\begin{abstract} 
\leftskip 54.8pt
\rightskip 54.8pt
We propose a self-consistent scheme for the determination of the 
ground-state (GS) properties of interacting electrons in a magnetic field, 
and of systems whose GS's time-reversal-symmetry (TRS) is spontaneously 
broken. It is based on a newly-developed many-body perturbation theory 
that is valid, irrespective of the strength of correlation, provided the 
GS number densities $n_{\uparrow}({\bf r})$, $n_{\downarrow}({\bf r})$, 
and the {\sl total} paramagnetic particle flux density are pure-state 
non-interacting $v$-representable. Our approach can in particular be 
applied to (modulated) two-dimensional electron systems in the 
fractional quantum-Hall regime.  
\end{abstract}

\pacs{71.27.+a, 73.40.Hm, 31.15.Md, 31.15.Ew [Published in:
Solid State Commun. {\bf 104}, 227 - 231 (1997).]} ]
\narrowtext

Consider the following Hamiltonian that in the non-relativistic
limit governs the behaviour of a general system of interacting 
electrons (we employ the SI units):
\begin{eqnarray}
\label{e1}
\widehat{H} {=} & & \sum_{\sigma} \int d^3r\; 
\widehat{\psi}_{\sigma}^{\dag}({\bf r}) 
\left\{ {1\over 2 m_e} \left[ -i\hbar\mbox{\boldmath $\nabla$} 
+ e \mbox{\boldmath $A$}({\bf r}) \right]^2\right. \nonumber\\
& &+ \left. 
v({\bf r}) + {1\over 2} g\mu_B B({\bf r}) \sigma_z(\sigma) \right\}
\widehat{\psi}_{\sigma}({\bf r}) + \widehat{V},
\end{eqnarray}
with $\widehat{V} {=} (1/2) \sum_{\sigma,\sigma'}
\int d^3r d^3r'$ $\widehat{\psi}_{\sigma}^{\dag}({\bf r})
\widehat{\psi}_{\sigma'}^{\dag}({\bf r}')$ $v_{c}({\bf r}
-{\bf r}')$ $\widehat{\psi}_{\sigma'}({\bf r}')
\widehat{\psi}_{\sigma}({\bf r})$
the interaction part (the quantities in these expressions are defined in 
Ref.~\cite{r1}). Here for simplicity we have assumed that the magnetic 
field $\mbox{\boldmath $B$}({\bf r}) \equiv \mbox{\boldmath $\nabla$} 
\wedge \mbox{\boldmath $A$}({\bf r})$, with $\mbox{\boldmath $A$}$ the 
external vector potential, points in a definite direction, i.e.
$\mbox{\boldmath $B$}({\bf r}) = B({\bf r}) \mbox{\boldmath $e$}_z$. 

Efforts for determining the electron-electron interaction effects on 
the physical properties of systems described by $\widehat{H}$ include, 
e.g., use of the many-body perturbation theory at some level of 
approximation, various quantum Monte-Carlo techniques and 
exact-diagonalisation methods. Application of the (ordinary) perturbation 
theory is considered to be limited to so-called `weakly-correlated' 
systems, and although suitable for dealing with `strongly-correlated' 
systems, the latter two categories of mentioned methods are applicable 
only to systems consisting of relatively small number of particles or those 
whose Hilbert space is small. In this Communication we introduce a {\sl 
self-consistent} many-body perturbation theory that, in principle, can 
be applied to weakly as well as strongly correlated systems. Here our 
attention will be mainly focussed towards some GS properties. In order 
to be able to describe GS's with spontaneously broken TRS,
we retain $\mbox{\boldmath $A$}$ in Eq.~(\ref{e1}) and identify 
the case of zero external magnetic field with $B({\bf r}) \to 0$.

In a previous work \cite{r2} we have discussed a fundamental problem from 
which any perturbation theory {\sl can} suffer: that despite the possible
convergence of a perturbation series, the ultimate results may not even 
approximately be related to the quantities of interest. The reason for this 
type of breakdown of the perturbation theory lies in that the GS of the 
non-interaction Hamiltonian $\widehat{H}_0 {=} \widehat{H} -\widehat{V}$ 
(possibly modified by some effective one-body term), may not be 
adiabatically connected with that of the fully interacting system \cite{r2a}. 

Our analysis in Ref.~\cite{r2} shows that a perturbation theory based 
on a non-interacting Hamiltonian $\widehat{H}_0^{'}$ whose ground state 
$\vert\Phi_0^{'}\rangle$ satisfies $\langle \Phi_0^{'}\vert \widehat{{\cal 
O}}_i \vert\Phi_0^{'}\rangle$ $=\langle\Psi_0\vert \widehat{{\cal O}}_i 
\vert\Psi_0\rangle$ ${\equiv}{\cal O}_i$, $i=1,2, \dots, {\cal N}$, with 
$\vert \Psi_0\rangle$ the GS of the fully interacting system and 
$\{\widehat{{\cal O}}_i \vert i=1,2,\dots, {\cal N}\} {\equiv} S_{\cal N}$ a 
specified set of quantities that {\sl uniquely} determine the {\sl ground 
state} of the system (see following paragraph), is an {\sl unconditionally} 
valid perturbation theory \cite{r2b}. We observe that for ${\cal O}_i$ 
$=\langle\Phi_0^{'}\vert \widehat{{\cal O}}_i\vert\Phi_0^{'}\rangle$ 
to be satisfied it is necessary that ${\cal O}_i$ be pure-state 
non-interacting $v$-representable (for definition see Ref.~\cite{r3}). 

For the work that we present in this Communication we rely on a theorem 
due to Vignale and Rasolt (VR) \cite{r4} according to which the GS
of the Hamiltonian given in Eq.~(\ref{e1}) is a {\sl unique} 
functional of the number densities $n_{\sigma}({\bf r})$, $\sigma\in 
\{\uparrow, \downarrow\}$, and the {\sl total} paramagnetic particle 
flux density $\mbox{\boldmath $j$}_{p}({\bf r}) {=} \mbox{\boldmath 
$j$}_{p;\uparrow} ({\bf r}) + \mbox{\boldmath $j$}_{p;\downarrow}({\bf r})$. 
If the external vector potential $\mbox{\boldmath $A$}$ were spin 
dependent, denoted by $\mbox{\boldmath $A$}_{\sigma}$, then the GS would 
become \cite{r4} a unique functional of $n_{\sigma}({\bf r})$ and 
$\mbox{\boldmath $j$}_{\sigma}({\bf r})$ for $\sigma = \uparrow,\downarrow$. 
To keep our approach general, so that both $\mbox{\boldmath 
$j$}_{p;\uparrow}$ and  $\mbox{\boldmath $j$}_{p;\downarrow}$, pertaining 
to the {\sl interacting} system, can be calculated \cite{r4}, in what 
follows we formally assume to have a spin-dependent external vector 
potential; for the actual calculations we take $\mbox{\boldmath 
$A$}_{\uparrow}$ $\equiv$ $\mbox{\boldmath $A$}_{\downarrow}$ $\equiv$ 
$\mbox{\boldmath $A$}$. Thus for the present case we have $S_{\cal N}$ 
$= S_{4} \equiv$ $\{\widehat{n}_{\sigma},$ $\widehat{\mbox{\boldmath 
$j$}}_{p;\sigma}\vert$ $\sigma = \uparrow, \downarrow\}$ in which 
$\widehat{n}_{\sigma} ({\bf r})$ ${=}\widehat{\psi}_{\sigma}^{\dag}({\bf r})
\widehat{\psi}_{\sigma}({\bf r})$, and $\widehat{\mbox{\boldmath 
$j$}}_{p;\sigma}({\bf r})$ ${=}(-i\hbar/[2 m_e])$ 
$\{\widehat{\psi}_{\sigma}^{\dag}({\bf r})$ $[\mbox{\boldmath 
$\nabla$}\widehat{\psi}_{\sigma}({\bf r})]$ $- [\mbox{\boldmath 
$\nabla$}\widehat{\psi}_{\sigma}^{\dag}({\bf r})]$ 
$\widehat{\psi}_{\sigma}({\bf r})\}$. The Hamiltonian $\widehat{H}_0^{'}$ 
for the case at hand coincides with the Kohn-Sham (KS) Hamiltonian as 
introduced by VR \cite{r4}. We have $\widehat{H}_{KS} = \sum_{\sigma} 
\int d^3r\; 
\widehat{\psi}_{\sigma}^{\dag}({\bf r}) H_{KS;\sigma}({\bf r})
\widehat{\psi}_{\sigma}({\bf r})$, in which \cite{r5}
\begin{eqnarray}
\label{e2}
H_{KS;\sigma}({\bf r}) {\equiv} 
& & {1\over 2 m_e} \left( -i\hbar\mbox{\boldmath $\nabla$} + e 
[\mbox{\boldmath $A$}({\bf r}) + \mbox{\boldmath $A$}_{xc;\sigma}
({\bf r}) ]\right)^2 \nonumber \\  
& &-{{e^2}\over 2 m_e} \left( A_{xc;\sigma}^2({\bf r})
+ 2 \mbox{\boldmath $A$}({\bf r})\cdot\mbox{\boldmath $A$}_{xc;\sigma}
({\bf r})\right)\nonumber \\
+ v({\bf r}) & & + {1\over 2} g\mu_B B({\bf r}) \sigma_z(\sigma)
+ v_H({\bf r}) + v_{xc;\sigma}({\bf r}).
\end{eqnarray}

Let now $G_{\sigma}({\bf r}t,{\bf r}'t')$ denote the single-particle
Green function corresponding to $\widehat{H}$, and $G_{KS;\sigma}$
that corresponding to $\widehat{H}_{KS}$. Making use of the results 
\begin{equation}
\label{e3}
\begin{array}{ll}
n_{\sigma}({\bf r}) = -i G_{\sigma}({\bf r}t,{\bf r}t+0^+), \\ \\
\mbox{\boldmath $j$}_{p;\sigma}({\bf r})
= {{-\hbar}\over 2 m_e} \lim_{{\bf r}'\to {\bf r}}
\left(\mbox{\boldmath $\nabla$}-\mbox{\boldmath $\nabla$}'\right)
G_{\sigma}({\bf r}t,{\bf r}'t+0^+),
\end{array}
\end{equation}
which, provided $n_{\sigma}$ and $\mbox{\boldmath $j$}_{p;\sigma}$ are
pure-state non-interacting $v$-representable, {\sl are} by construction 
also valid for $G_{\sigma} \to G_{KS:\sigma}$, and a perturbation expansion 
for $G_{\sigma}$ in terms of $G_{KS;\sigma}$, we obtain for a $d$-dimensional 
system a set of $2(d+1)$ coupled non-linear equations which we have 
diagrammatically represented in Fig.~1. From these equations the two 
$(d+1)$-vectors $(v_{xc;\sigma},\mbox{\boldmath $A$}_{xc;\sigma})$, 
$\sigma \in \{\uparrow,\downarrow\}$, can be determined and consequently 
the $\widehat{H}_{KS}$ on the basis of which an unconditionally valid
perturbation theory can be set up \cite{r2}. The unfamiliar Feynman 
diagrams in Fig.~1 have their origin in the specific choice for the 
``unperturbed'' Hamiltonian. For $\widehat{H}_1 \equiv \widehat{H} - 
\widehat{H}_{KS}$ we have
\begin{eqnarray}
\label{e4}
\widehat{H}_1 & & = -e \sum_{\sigma} \int d^3r\; \mbox{\boldmath $A$}_{xc;
\sigma}({\bf r})\cdot \widehat{\mbox{\boldmath $j$}}_{p;\sigma}({\bf r})
\nonumber\\
- & & \sum_{\sigma} \int d^3r\; 
\widehat{\psi}_{\sigma}^{\dag}({\bf r})
\left\{v_{H}({\bf r}) + v_{xc;\sigma}({\bf r})\right\} 
\widehat{\psi}_{\sigma}({\bf r})
+ \widehat{V}.
\end{eqnarray}
For casting $\widehat{H}_1$ into a form that makes application of 
the standard procedures of the many-body perturbation theory \cite{r6} 
possible, we define the following operator-valued {\sl non-local} 
potential:
\begin{equation}
\label{e5}
u_{xc;\sigma}({\bf r},{\bf r}') {\equiv}
{{i e\hbar}\over 2 m_e} \delta({\bf r}-{\bf r}')
\mbox{\boldmath $A$}_{xc;\sigma}({\bf r})\cdot
\left(\mbox{\boldmath $\nabla$} -\mbox{\boldmath $\nabla$}'\right).
\end{equation}
Through this the first term on the right-hand side of Eq.~(\ref{e4})
can be written as $[- \sum_{\sigma}$ $ \int d^3r d^3r'$ $u_{xc;\sigma}
({\bf r},{\bf r}')$ $\widehat{\psi}_{\sigma}^{\dag}({\bf r})
\widehat{\psi}_{\sigma}({\bf r}')]$. In Fig.~1 the directed wiggly
lines, pointing from ${\bf r}'\sigma$ to ${\bf r}\sigma$, stand for 
$(-i/\hbar)[-u_{xc;\sigma}({\bf r},{\bf r}')]$. For the rules concerning 
evaluation of contribution of diagrams see Ref.~\cite{r6}. 

Earlier Sham \cite{r6a}, by imposing $G=G_{KS}$ at ${\bf r}'= {\bf r}$, 
$t'=t+0^+$ for systems of spin-less electrons (with no broken TRS), 
obtained an implicit expression for $v_{xc}$. Sham, however, did not 
address the problem concerning the validity of the many-body perturbation 
theory which is central to our present work.

Let the contributions of diagrams $(a_1)$, $(b_1)$ and $(c_1)$ in 
Fig.~1 be denoted by $n_{\sigma}^{(x)}({\bf r})$ and $\mbox{\boldmath 
$j$}_{p;\sigma}^{(x)}({\bf r})$, with $x= a_1, b_1, c_1$. The
non-linear equations associated with these first-order diagrams read (with 
$\sigma \in \{\uparrow ,\downarrow\}$)
\begin{figure}[t!]
\protect
\label{f1}
\centerline{ 
\psfig{figure=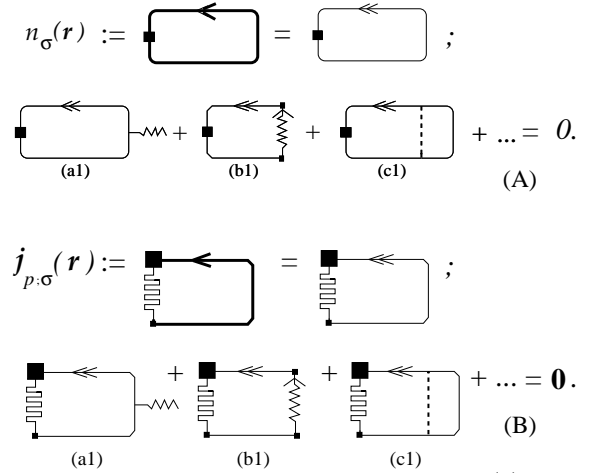,width=3.00in} }
\caption{
Diagrammatic representation of Eq.~(\protect\ref{e3}) with $G_{\sigma}$ 
(thick line with single arrow) and $G_{KS;\sigma}$ (double-arrowed thin 
line) on the right-hand sides (upper parts of (A) and (B)). Contributions 
of diagrams $(a_1)$, $(b_1)$, $(c_1)$, $\dots$ add to zero for the correct 
$v_{xc;\sigma}$ (wiggly line with one loose end; the associated expression 
equals $(-i/\hbar)[-v_{xc;\sigma}]$) and ${\bf A}_{xc;\sigma}$ which defines 
$u_{xc;\sigma}$ (directed wiggly line) according to Eq.~(\protect\ref{e5}). 
The pulse-train-like line in (B) implies the non-local operation 
$(-\hbar/[2m_e])$ $\lim_{{\bf r}'\to {\bf r}}$ $(\nabla - \nabla')$ in 
Eq.~(\protect\ref{e3}). Lower parts of (A) and (B) correspond, respectively,
to the first and the second equation in Eq.~(\protect\ref{e6}). Note that 
in first order only one diagram (i.e. $(c_1)$) {\sl explicitly} depends 
on the Coulomb interaction, represented by the broken line (corresponding to 
$(-i/\hbar)v_c$); this line can be taken to represent also the dynamically 
screened interaction function.
}
\end{figure}
\begin{equation}
\label{e6}
\begin{array}{ll}
 n_{\sigma}^{(a_1)}({\bf r})
+n_{\sigma}^{(b_1)}({\bf r})
+n_{\sigma}^{(c_1)}({\bf r}) = 0,\\ \\
 \mbox{\boldmath $j$}_{p;\sigma}^{(a_1)}({\bf r})
+\mbox{\boldmath $j$}_{p;\sigma}^{(b_1)}({\bf r})
+\mbox{\boldmath $j$}_{p;\sigma}^{(c_1)}({\bf r}) = {\bf 0},
\end{array}
\end{equation}
in which
\begin{equation}
\label{e7}
\begin{array}{ll}
n_{\sigma}^{(x)}({\bf r}) \equiv - 2 {\rm Re}
\left\{\sum_s^{>} \sum_{s'}^{<} 
U_{\sigma;s,s'}^{(x)}\varrho_{\sigma;s',s}({\bf r}) \right\},\\ \\
\mbox{\boldmath $j$}_{p;\sigma}^{(x)}({\bf r}) \equiv 
-2{\rm Re}\left\{\sum_s^{>} \sum_{s'}^{<}
U_{\sigma;s,s'}^{(x)}
\mbox{\boldmath $J$}_{\sigma;s',s}({\bf r})\right\};
\end{array}
\end{equation}
\begin{equation}
\label{e8}
\begin{array}{ll}
U_{\sigma;s,s'}^{(a_1)} {\equiv} \int d^3r'\; v_{xc;\sigma}({\bf r}')
\varrho_{\sigma;s,s'}({\bf r}')/
(\varepsilon_{\sigma;s'} - \varepsilon_{\sigma;s}),\\ \\
U_{\sigma;s,s'}^{(b_1)} {\equiv} e \int d^3r'\; 
\mbox{\boldmath $A$}_{xc;\sigma}({\bf r}')
\cdot \mbox{\boldmath $J$}_{\sigma;s,s'}({\bf r}')/
(\varepsilon_{\sigma;s'} - \varepsilon_{\sigma;s}),\\ \\
U_{\sigma;s,s'}^{(c_1)} 
{\equiv} \sum_{s''}^{<} \int d^3r'\; 
\varrho_{\sigma;s,s''}({\bf r}') \\ \\
\times \left[\int d^3r''\; 
v_c({\bf r}'-{\bf r}'') 
\varrho_{\sigma;s'',s'}({\bf r}'') \right] /
(\varepsilon_{\sigma;s'} - \varepsilon_{\sigma;s}), \\ \\
\varrho_{\sigma;s,s'}({\bf r}){\equiv} 
\psi_{\sigma;s}^*({\bf r})\psi_{\sigma;s'}({\bf r});\;\;
\mbox{\boldmath $J$}_{\sigma;s,s'}({\bf r}){\equiv}(-i\hbar/[2 m_e]) \\ \\
\;\;\;\times\{\psi_{\sigma;s}^*({\bf r})[\mbox{\boldmath $\nabla$}
\psi_{\sigma;s'}({\bf r})] - [\mbox{\boldmath $\nabla$}
\psi_{\sigma;s}^*({\bf r})] \psi_{\sigma;s'}({\bf r})\}.
\end{array}
\end{equation}
We have $n_{\sigma}({\bf r}) = \sum_s^{<}\varrho_{\sigma;s,s}({\bf r})$, 
$\mbox{\boldmath $j$}_{p;\sigma}({\bf r}) = \sum_s^{<} \mbox{\boldmath 
$J$}_{\sigma;s,s}({\bf r})$. For $\mbox{\boldmath $A$}\equiv
\mbox{\boldmath $A$}_{xc}\equiv {\bf 0}$, the first equation in 
Eq.~(\ref{e6}) reduces to one embodying the `optimised-potential 
method' \cite{r6b}.

Above $\sum_{s}^{<}$ $(\sum_{s}^{>})$ stands for $\sum_{s=1}^{N_{\sigma}}$ 
$(\sum_{s=N_{\sigma}+1}^{\infty})$ and $\psi_{\sigma;s}$ denotes an 
eigenfunction of $H_{KS;\sigma}$ in Eq.~(\ref{e2}), with 
$\varepsilon_{\sigma;s}$ the corresponding eigenvalue 
($\varepsilon_{\sigma;s}$ $\leq$ $\varepsilon_{\sigma;s+1}$). Thus, 
Eq.~(\ref{e6}) not only explicitly depends on $v_{xc;\sigma}$ and 
$\mbox{\boldmath $A$}_{xc;\sigma}$ (via Eqs.~(\ref{e7}), (\ref{e8})), 
but also implicitly, through $\{\psi_{\sigma;s},\varepsilon_{\sigma;s}\}$. 
Since in solving Eq.~(\ref{e6}) we exactly diagonalise the KS Hamiltonian 
corresponding to any $(v_{xc;\sigma}, \mbox{\boldmath $A$}_{xc;\sigma})$, 
the self-consistent $(v_{xc;\sigma}, \mbox{\boldmath $A$}_{xc;\sigma})$ 
takes account of the electron-electron interaction effects, in so far as 
present in the employed diagrammatic expansion for $G_{\sigma}$, to 
infinite order. Moreover, it can be shown (cf. Ref.~\cite{r2}) that a 
solution $(v_{xc;\sigma}, \mbox{\boldmath $A$}_{xc;\sigma})$ of 
Eq.~(\ref{e6}) based on a finite-order perturbation expansion for 
$G_{\sigma}$ annihilates the combined contributions to $n_{\sigma}$ and 
$\mbox{\boldmath $j$}_{p;\sigma}$ of {\sl all} higher-order diagrams 
that have a common part in addition to one of the lower-order 
number-density diagrams (Fig.~1 (A)) that have been taken into account. 
Note that in solving Eq.~(\ref{e6}), $N_{\uparrow}$ and $N_{\downarrow}$, 
satisfying $N_{\uparrow} + N_{\downarrow} = N_e$, must be calculated 
self-consistently: for a given $N_e$, $N_{\uparrow}$, or $N_{\downarrow}$, 
is determined by the requirement that $\varepsilon_{\uparrow;1}, \dots, 
\varepsilon_{\uparrow; N_{\uparrow}}$, $\varepsilon_{\downarrow;1}, \dots, 
\varepsilon_{\downarrow; N_{\downarrow}}$ are the {\sl lowest} $N_e$ 
eigenvalues of $\{H_{KS;\sigma}\}$.

The non-linearity of Eq.~(\ref{e6}) implies that there is a multiplicity 
of solutions. The uniqueness theorem of VR \cite{r4} implies, however, 
that only one of the solutions corresponds to the GS (solutions 
$\alpha+v_{xc;\uparrow}$, $\alpha+v_{xc;\downarrow}$, with $\alpha$ a 
constant, are identical). It is important that {\sl all} solutions of 
Eq.~(\ref{e6}) correspond to {\sl some} \cite{r2} eigenstate of 
$\widehat{H}$. To single out the GS solution $(v_{xc;\sigma},\mbox{\boldmath 
$A$}_{xc;\sigma})$, let the GS of the system corresponding to the external 
potential $v=v_0$ and the external magnetic field strength $B=B_0$ (and 
therefore $\mbox{\boldmath $A$}_0$, in some gauge) be known. For the actual 
potential $v$ and magnetic field strength $B$ we define: $v^{(\lambda)}$ 
${=}v_0$ $+\lambda (v - v_0)$ and  $\mbox{\boldmath $A$}^{(\lambda')}$ 
${=}\mbox{\boldmath $A$}_0$ $+\lambda' (\mbox{\boldmath $A$} - 
\mbox{\boldmath $A$}_0)$. For determining the GS solution of the non-linear 
equations we choose some trajectory on the $(\lambda,\lambda')$-plane 
connecting $(0,0)$ with $(1,1)$. By starting from $(0,0)$, at small steps 
along the trajectory we solve Eq.~(\ref{e6}) to self-consistency. In doing 
so, at each step we take the SC solution of the immediately earlier step 
as the trial solution. In the event of encountering degeneracy (level crossing 
at, say, $(\lambda_0,{\lambda'}_0)$), we can easily select out the GS 
solution by realizing that at the point of degeneracy the energies of the 
degenerate states have discontinuous derivatives and the jump in the 
derivative of the GS total-energy curve is {\sl always} negative and has 
the largest {\sl magnitude} amongst the jump values corresponding to 
other energy curves. Now since the solutions found pertain to {\sl 
eigenstates} of the Hamiltonian corresponding to $v^{(\lambda = 
{\lambda}_0)}$ and $\mbox{\boldmath $A$}^{(\lambda' = {\lambda'}_0)}$, 
we can apply the Hellmann-Feynman theorem \cite{r6}, \cite{r2} which 
provides us with the derivatives of the eigenenergies with respect to 
$\lambda$ and $\lambda'$. A simple calculation shows that 
$n_{\uparrow}^{(\lambda)}$, $n_{\downarrow}^{(\lambda)}$ and 
$\mbox{\boldmath $j$}_{p;\uparrow}^{(\lambda')} + \mbox{\boldmath 
$j$}_{p;\downarrow}^{(\lambda')}$ are sufficient for determining the 
mentioned derivatives \cite{r7}. Note that through integration of the 
available total-energy derivatives along the chosen path, the GS total 
energy is readily obtained.
\begin{figure}[t!]
\protect
\label{f2}
\centerline{ 
\psfig{figure=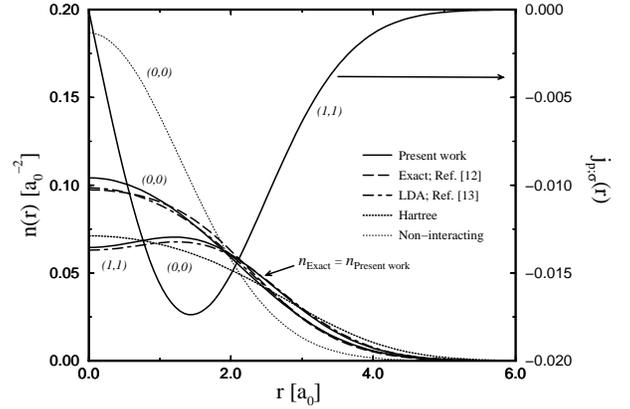,width=3.00in} }
\caption{Particle number and flux densities corresponding to a 
two-electron quantum dot in the $(M_z,S_z)$ state with $M_z$ and 
$S_z$, respectively, the orbital- and spin-angular momentum quantum 
numbers along the quantisation axis. The parameters chosen are those 
employed in Refs.~\protect\cite{r8},\protect\cite{r9}: $B=1.0$ T, $v(r) 
= m_e^* \Omega^2 r^2/2$ with $m_e^* = 0.067 m_e$ and $\Omega=3.37$ meV; 
$\epsilon_r =12.4$, $g=-0.44$; see Ref.~\protect\cite{r1}; the effective 
Bohr radius $a_0$ equals $9.794\times 10^{-9}$ m. The current density 
$-e j_{p;\sigma}(r)$, with $\sigma =\uparrow$, is in units of $2.95\times 
10^2$ A/m (note the $-e$); $j_{p;\sigma}$, the azimuthal component of 
${\bf j}_{p;\sigma}$ (the radial component is vanishing), corresponds to 
the symmetric gauge, ${\bf A} = {\bf B}\wedge {\bf r}/2$. } 
\end{figure}
Our proposed framework yields $n_{\sigma}$ and $\mbox{\boldmath
$j$}_{p;\sigma}$ that exactly satisfy the static equation of 
continuity. This is because both of these are derived from the same 
single-particle Green function, $G_{KS;\sigma}$. Further, provided the 
`associated' diagrams (see further on) corresponding to a particular 
order of the perturbation expansion for $G_{\sigma}$ are taken into 
account, our framework is also gauge invariant. To see this clearly, 
let the $4$-potential $(v_{xc;\sigma},\mbox{\boldmath $A$}_{xc;\sigma})$ 
correspond to $(v,\mbox{\boldmath $A$}_{\sigma})$. For definiteness 
suppose we have obtained the former by taking into account diagrams 
$(a_1)$, $(b_1)$ and $(c_1)$ in Fig.~1, or by solving Eq.~(\ref{e6}). 
Since $\mbox{\boldmath $A$}_{\sigma} \to$ ${\mbox{\boldmath 
$A$}'}_{\sigma} \equiv \mbox{\boldmath $A$}_{\sigma} + \mbox{\boldmath 
$\nabla$}{\Lambda}_{\sigma}$ in a gauge-invariant theory must lead to 
$n_{\sigma} \to n_{\sigma}$ and $\mbox{\boldmath $j$}_{p;\sigma} \to 
\mbox{\boldmath $j$}_{p;\sigma} - (e/m_e) n_{\sigma} \mbox{\boldmath 
$\nabla$} {\Lambda}_{\sigma}$, it follows that for $v_{xc;\sigma}'$ and 
$\mbox{\boldmath $A$}_{xc;\sigma}'$, corresponding to ${\mbox{\boldmath 
$A$}'}_{\sigma}$, hold \cite{r4}: $v_{xc;\sigma}' \equiv v_{xc;\sigma}$ 
$+ (e^2/m_e)$ $\mbox{\boldmath $A$}_{xc;\sigma}$ $\cdot \mbox{\boldmath 
$\nabla$}$ ${\Lambda}_{\sigma}$ and $\mbox{\boldmath $A$}_{xc;\sigma}'$ 
$\equiv\mbox{\boldmath $A$}_{xc;\sigma}$. Our framework would be gauge 
non-invariant if Eq.~(\ref{e6}) would not be satisfied by $(v_{xc;\sigma}',
\mbox{\boldmath $A$}_{xc;\sigma}')$.  We now show that this is not the 
case. First, $\varepsilon_{\sigma;s}$ does not depend on the choice of 
gauge. Further, from the explicit expressions in Eq.~(\ref{e8}) it can 
be shown that $\varrho_{\sigma;s,s'}$, similar to $n_{\sigma}$, is gauge 
invariant, and that for $\mbox{\boldmath $A$}_{\sigma} \to {\mbox{\boldmath 
$A$}'}_{\sigma}$, $\mbox{\boldmath $J$}_{\sigma;s,s'}$ $\to$ 
$\mbox{\boldmath $J$}_{\sigma;s,s'}$ $- (e/m_e) \varrho_{\sigma;s,s'}$ 
$\mbox{\boldmath $\nabla$} \Lambda_{\sigma}$. It follows from 
Eq.~(\ref{e8}) that $U_{\sigma;s,s'}^{(a_1)}$ $+ U_{\sigma;s,s'}^{(b_1)}$ 
and $U_{\sigma;s,s'}^{(c_1)}$ are gauge invariant (diagrams $(a_1)$ and 
$(b_1)$ in Fig.~1 are `associated'). Thus $n_{\sigma}^{(a_1)}$ $+ 
n_{\sigma}^{(b_1)}$ and $n_{\sigma}^{(c_1)}$ are gauge invariant and 
$\mbox{\boldmath $j$}_{p;\sigma}^{(a_1)}$ $+ \mbox{\boldmath 
$j$}_{p;\sigma}^{(b_1)}$ $\to$ $(\mbox{\boldmath $j$}_{p;\sigma}^{(a_1)}$ 
$+ \mbox{\boldmath $j$}_{p;\sigma}^{(b_1)})$ $- (e/m_e)$ 
$(n_{\sigma}^{(a_1)}$ $+ n_{\sigma}^{(b_1)})$ $\mbox{\boldmath $\nabla$}$ 
$\Lambda_{\sigma}$, $\mbox{\boldmath $j$}_{p;\sigma}^{(c_1)}$ $\to$ 
$\mbox{\boldmath $j$}_{p;\sigma}^{(c_1)}$ $- (e/m_e)$ $n_{\sigma}^{(c_1)}$ 
$\mbox{\boldmath $\nabla$}$ $\Lambda_{\sigma}$. Therefore satisfaction of 
Eq.~(\ref{e6}) does not depend on the choice of gauge.

We have applied our formalism to a cylindrically symmetric quantum dot 
with a parabolic confining potential, taking into account only the 
first-order diagrams that are explicitly shown in Fig.~1. In Fig.~2 we 
present the calculated electronic number and flux densities in the GS's
of definite symmetries (see caption) and compare the former with some 
available results \cite{r8},\cite{r9} --- note that 
$\mbox{\boldmath $j$}_{p;\sigma} \equiv {\bf 0}$ in the $(M_z=0,S_z=0)$ 
GS. It turns out that in the $(1,1)$ state, our calculated $n$ $[\equiv 
n_{\uparrow} + n_{\downarrow}$; for the $(1,1)$ state, $n=n_{\uparrow}]$ 
is, within the numerical accuracy of the calculations, identical with the 
{\sl exact} $n$. On the other hand, for small values of $r$ our 
$(0,0)$ GS $n(r)$ overestimates the exact $n(r)$ by several percents. 
Both of these density profiles are almost identical with the Hartree-Fock 
(HF) results \cite{r8}. There is however a fundamental difference between 
the HF results and those according to the present scheme. According to 
the HF approach, for all $B\not=0$, the GS of {\sl the system under 
consideration} is a $(1,1)$ state, in obvious contradiction with the 
exact results \cite{r8}. For instance, for $B=1$T, the exact GS is a 
$(0,0)$ state, and the (first) transition to the $(1,1)$ state takes 
place at $B=2.0$T. In agreement with this, we find for $B=1$T the 
lowest-lying state to be a $(0,0)$ state. It should be mentioned that 
the results in Fig.~2 labelled by $(1,1)$ correspond to the {\sl 
lowest-energy} $(1,1)$ state; for $B=1$T, as indicated, the state 
corresponding to the {\sl absolute} minimum of energy is a $(0,0)$ 
state. It is therefore important to emphasise that the VR theorem 
\cite{r4} is also valid for excited states which are minimum-energy 
states corresponding to specified symmetries. This follows from the fact 
that the variational principle, which underlies the VR theorem, can also 
be applied in symmetry-restricted Hilbert spaces. For comparison, in 
Fig.~2 we present the results obtained within the 
local-density-approximation (LDA) scheme \cite{r9}. 

Concerning the overestimation in the vicinity of $r=0$ of our calculated 
$n(r)$ corresponding to the $(0,0)$ state, our preliminary calculations 
indicate that this is substantially suppressed through replacing $v_c$ in 
diagrams $(c1)$ of Fig.~1 by the dynamically-screened interaction 
function within the random-phase approximation. 

We have analysed the asymptotic behaviour of the functions
$v_{xc;\sigma}({\bf r})$, $\mbox{\boldmath $A$}_{xc;\sigma}({\bf r})$,
for $\vert{\bf r}\vert\to 0,\infty$, corresponding to the system under 
consideration, both within the framework of the present formalism and 
that of the LDA. The results will be reported elsewhere \cite{r7}. We 
only mention that unless appropriate measures are taken, the 
current-carrying GS's of the LDA \cite{r4} are unstable.

In conclusion, we have introduced a self-consistent perturbation theory 
for interacting electrons in presence or absence of an external magnetic 
field. In the latter case the system can possibly have a spontaneously 
broken TRS, such as is the case in open-shell atoms.
Already the first stage in the application of this theory provides us 
with the scalar and vector exchange-correlation potentials that determine
the Kohn-Sham Hamiltonian within the framework of the current-and-spin 
density-functional theory. This Hamiltonian forms the basis for 
construction of reliable perturbation expansions for various quantities, 
including those corresponding to the excited states of the interacting 
system, such as energies of the elementary excitations. We propose use 
of our method for determining properties of (modulated) two-dimensional 
electron systems in the fractional quantum-Hall regime where electrons 
are strongly correlated. Work in this direction is in progress.

{\small
The author should like to thank Professors Lars Hedin, Rolf 
Gerhardts, Vidar Gudmundsson, Drs Andrei Manolescu, Daniela 
Pfannkuche, and Professor Dieter Weiss for helpful discussions. 
The support of Max-Planck-Gesellschaft is gratefully acknowledged.}


\begin{references}
\bibitem{r1}
In Eq.~(\ref{e1}), $\widehat{\psi}_{\sigma}^{\dag}$ and 
$\widehat{\psi}_{\sigma}$, $\sigma\in\{\uparrow,\downarrow\}$, are 
creation and annihilation field operators; $g$ is the Land\'e factor, 
$v$ the external potential (e.g., the confining potential), $\mu_B$ 
the Bohr magneton; $\sigma_z(\sigma) =\pm 1$ for $\sigma= 
\uparrow,\downarrow$; $v_c({\bf r}-{\bf r}') {\equiv} 
e^2/(4\pi\epsilon_0\epsilon_r \vert {\bf r}-{\bf r}'\vert)$, with 
$-e$ $(<0)$ the electron charge, $\epsilon_0$ the vacuum permittivity 
and $\epsilon_r$ the relative dielectric constant of the system. For 
band electrons, the electron mass $m_e$ should be replaced by the 
effective mass $m_e^*$.
\bibitem{r2b}
Suppose $\vert\Phi_0^{'}\rangle$ were not adiabatically connected with 
$\vert\Psi_0\rangle$. Owing to the $v$-representability of $\{{\cal O}_i\}$, 
it follows that the adiabatically-evolved $\vert\Phi_0^{'}\rangle$ is 
the {\sl GS} of {\sl some} many-body $\widehat{H}'$ that differs from 
$\widehat{H}$ by non-trivial one-body terms (e.g., for $\widehat {H}$ 
in Eq.~(\ref{e1}), $\widehat{H}'$ would involve $v'$ and ${\bf A}'$, 
with, e.g., $v'$ differing by more than a constant from $v$). This 
contradicts the fact that {\sl GS}'s of such Hamiltonians {\sl cannot} 
\cite{r3} have identical $\{{\cal O}_i\}$.
\end{references}
\end{document}